\documentclass[letterpaper,twocolumn,english]{revtex4-1}
\usepackage{ae,aecompl}
\usepackage{helvet}

\usepackage[T1]{fontenc}
\usepackage[latin9]{inputenc}
\usepackage{mathrsfs}
\usepackage{amsmath}
\usepackage{amssymb}
\usepackage{graphicx}



\usepackage{babel}
\begin{document}
\title{Josephson Ladders as a Model System for 1D Quantum Phase Transitions}
\author{M.T. Bell$^{1}$, B. Dou\c{c}ot$^{2}$, M.E. Gershenson$^{3}$, L.B.
Ioffe$^{2}$, and A. Petkovi\'{c}$^{4}$ }
\affiliation{$^{1}$Department of Electrical Engineering, University of Massachusetts
Boston, Boston, MA 02125, USA }
\affiliation{$^{2}$LPTHE, Sorbonne Université and CNRS, Paris, France}
\affiliation{\textrm{$^{3}$}Department of Physics and Astronomy, Rutgers University,
Piscataway, NJ 08854, USA}
\affiliation{$^{4}$Laboratoire de Physique Théorique, Université de Toulouse,
CNRS, UPS, France}
\begin{abstract}
We propose a novel platform for the study of quantum phase transitions
in one dimension (1D QPT). The system consists of a specially designed
chain of asymmetric SQUIDs; each SQUID contains several Josephson
junctions with one junction shared between the nearest-neighbor SQUIDs.We
develop the theoretical description of the low energy part of the
spectrum. In particular, we show that the system exhibits the quantum
phase transition of Ising type. In the vicinity of the transition
the low energy excitations of the system can be described by Majorana
fermions. This allow us to compute the matrix elements of the physical
perturbations in the low energy sector. In the microwave experiments
with this system, we explored the phase boundaries between the ordered
and disordered phases and the critical behavior of the system's low-energy
modes close to the transition. Due to the flexible chain design and
control of the parameters of individual Josephson junctions, future
experiments will be able to address the effects of non-integrability
and disorder on the 1D QPT.
\end{abstract}
\maketitle

\section{Introduction}

The idea of quantum simulations emerged in the early '80s out of the
realization of the fundamental difficulty of emulating complex quantum
system using classical computers. Over time, this idea has evolved
into a sub-field of quantum computing \cite{QSimulations}. Similar
to the quantum computers, quantum simulators are based on the networks
of quantum bits (qubits), but, in contrast to the fully-fledged quantum
computers, quantum simulators do not employ discrete gate operations
and error correction codes. The quantum simulators with tunable parameters
are designed to emulate only certain types of Hamiltonians. However,
it is possible to show that a very general Hamiltonian can be simulated
by seemingly restricted class of spin chains with XX and YY interactions
\cite{Cubitt2016,Cubitt2017}. One hopes that such simulators will
facilitate designing novel quantum systems and exploration of phenomena
and regimes inaccessible in the past. Furthermore, quantum simulators
enable the experimental study of quantum annealing in the context
of adiabatic quantum computation. Modern quantum simulators are based
on several platforms, which include cold atoms \cite{coldatoms2012,coldatom2017},
cold ions \cite{coldions2012}, and superconducting qubits \cite{houck2012}. 

Our research focuses on designing artificial spin systems - tunable
1D Josephson arrays with controlled interactions \textendash{} that
emulate the quantum 1D models. The integrable model of a 1D Ising
spin chain in the transverse magnetic field serves as a paradigm in
the context of nonequilibrium thermodynamics and quantum critical
phenomena \cite{Dutta2015,Sachdev1999}. Both the transverse field
Ising and XY models, being relevant to a broad range of physical systems,
played a crucial role in the understanding of quantum phase transitions
\cite{Sachdev1999}. These models have generated a formidable body
of the theoretical activity over the past fifty years. Recently, these
models played a crucial role in the development of quantum annealing
techniques and adiabatic quantum algorithms \cite{Johnson2011}. 

In the past, the experimental study of quantum spin dynamics in 1D
has been largely limited to microscopic spins in condensed-matter
systems. It has been demonstrated that such quasi-1D spin materials
as LiHoF$_{4}$ and CoNb$_{2}$O$_{6}$ can be continuously tuned
across the quantum phase transition (QPT) \cite{Bitko1996,Ronnow2007,Coldea2010}.
Though these works opened up new vistas in the studies of transverse
field Ising model, the experimental realization of 1D quantum spin
models in well-controllable and tunable systems remains a challenge.
Indeed, as an experimental tool, the quasi-1D spin systems in solids
are limited in several respects: (i) the inter-chain interactions
are not negligibly weak, and, thus, these systems are inevitably quasi-1D,
(ii) the exchange interactions between the nearest-neighbor spins
cannot be varied, (iii) the exchange interactions are the same for
all pair of spins, which does not allow for exploring the effect of
disorder and phase boundaries without adding a significant amount
of impurities, and (iv) the available experimental tool for these
systems - scattering of neutrons - interacts only with a narrow class
of excitations. Flexibility in the design of artificial spin systems,
which are free from these limitations, facilitates bridging the gap
between the theoretical study of ideal spin chains and the experimental
investigation of bulk magnetic samples. In particular, this flexibility
allows one to address an important issue of the effects of disorder
on the statics and dynamics of transverse field spin models. Recently,
the transverse-field Ising model was realized in the chain of artificial
and fully-controllable spins - eight flux qubits with tunable spin\textendash spin
couplings \cite{Johnson2011}. We pursue a similar approach using
specially designed one-dimensional Josephson ladders.

The paper is organized as follows. In Section II we introduce the
Josephson ladder as a novel platform for the study of the 1D quantum
phase transitions. The theory of these quantum systems is presented
in Section III. The microwave experiments on characterization of the
spectra of these novel quantum systems are described in Section IV.
The outlook is provided in Section V.

\section{1D Josephson Ladders}

\begin{figure*}
\includegraphics[width=1\textwidth]{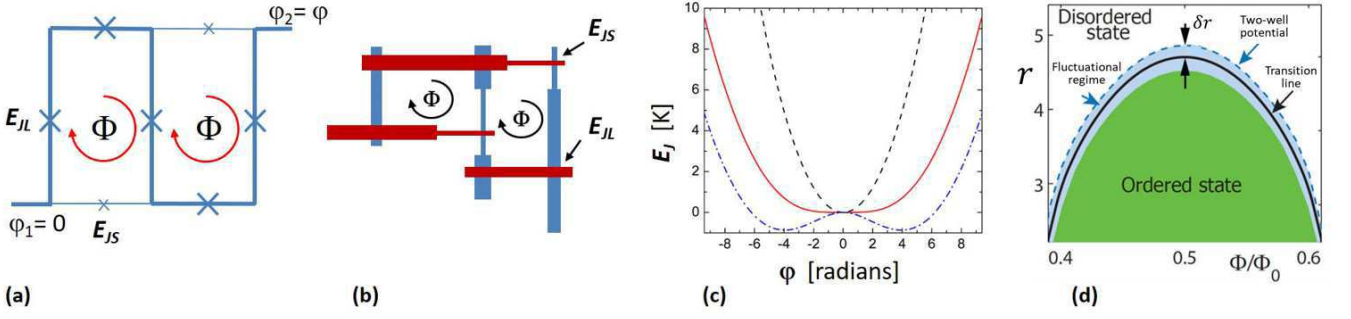}

\caption{The unit cells, potential energy, and the phase diagram of the Josephson
ladder \cite{Bell2012}. Panel (a): The unit cells of the ladder include
smaller and larger Josephson junctions with Josephson energies $E_{JS}$
and $E_{JL}$, respectively. All cells are threaded by the same magnetic
flux $\Phi$; the phase difference across the ladder is $\varphi_{2}-\varphi_{1}=\varphi$.
Panel (b): The design of the ladder. Bottom electrodes of the Josephson
tunnel junctions are shown in blue, top electrodes - in red. Panel
(c): The Josephson energy $E_{J}(\varphi)$ of a ladder with the ratio
$E_{JL}/E_{JS}\equiv r<r_{0}$ calculated for three values of the
flux: $\Phi=0$ (dashed curve), $\Phi=\Phi_{C}$ (solid curve), and
$\Phi=\Phi_{0}/2$ (dash-dotted curve). The energy is periodic in
phase with the period $2\pi,$ here we show only a single branch.
Panel (d): The phase diagram of the ladders on the $r$-plane. For
$r<r_{c}(\Phi)$ the broken-symmetry phase is formed at $\Phi_{C}<\Phi<\Phi_{0}-\Phi_{C}$
as a result of the phase transition of the Ising type. The boundary
of this ordered-state region is shown as the solid black curve. The
extension of the fluctuational regime is of the order of $\delta r\approx0.1$
for realistic arrays. \label{fig:elementary-cell}}
\end{figure*}

The Josephson ladders designed for the study of the 1D QPT represent
a 1D chain of coupled asymmetric SQUIDs (Fig. \ref{fig:elementary-cell}).
The unit cell of the ladders is similar to that of the Josephson arrays
developed as superinductors (the elements with the microwave impedance
much greater than the resistance quantum $h/e^{2}$) \cite{Bell2012}.
Each unit cell contains a single smaller junction with the Josephson
energy $E_{JS}$ in one arm and three larger Josephson junctions with
the Josephson energy $E_{JL}$ in the other arm (Fig. \ref{fig:elementary-cell}a).
The adjacent cells are coupled via one larger junction. 

The ladders are characterized by the ratio of Josephson energies of
the larger and smaller junctions, $r\equiv E_{JL}/E_{JS}$. For $r$
greater than the critical value $r_{0}$ the Josephson energy of the
ladder as a function of the phase $\varphi$ across the ladder has
only one minimum at $\varphi=0$ regardless of the magnitude of the
external magnetic field perpendicular to the ladder\textquoteright s
plane. The value of $r_{0}$ depends on the ladder length and the
strength of quantum fluctuations (in the quasiclassical case, $r_{0}=5$
for an infinitely long ladder) (see Section III). The regime $r>r_{0}$,
where the ladder can be characterized by the Josephson inductance
$L_{J}\propto(\frac{d^{2}E_{J}(\varphi)}{d\varphi^{2}})^{-1}$ \cite{tinkham2004},
was comprehensively studied in \cite{Bell2012}. 

If $r<r_{0}$, the double-minima dependence $E_{J}(\varphi)$ is realized
over a range of the magnetic flux $\Phi_{C}<\Phi<\Phi_{0}-\Phi_{C}$
where $\Phi_{C}(r)$ is the critical flux, $\Phi_{0}=\pi\hbar/e$
is the flux quantum. Figures \ref{fig:elementary-cell}c,d show that
at a fixed $r<r_{0}$, the spontaneous symmetry breaking occurs at
a critical value of the flux $\Phi_{C}(r)$ and the single-minimum
energy $E_{J}(\varphi)$ is transformed into a double-minima function.
The Josephson inductance diverges with $\Phi$ approaching $\Phi_{C}(r)$
(hence, the term ``superinductance''); quantum phase fluctuations
eliminate the divergency. 

In the double-minima regime, the direction of currents in the ladder
unit cells induced by an external magnetic field can be viewed as
two states of the 1/2 pseudo-spins. As the potential barrier between
these states increases, the quantum phase fluctuations decrease and
the global broken-symmetry state emerges. The appearance of the double
well potential in a single cell does not imply a global phase transition
in the whole system. The large-scale quantum fluctuations destroy
the global order parameter if the barrier between two minima is too
small. The global order parameter is formed at $r_{c}(\Phi)<r_{0}$
as a result of the phase transition of the Ising type ($r_{c}(\Phi)$
corresponds to the solid black curve in Fig. \ref{fig:elementary-cell}d).
The extension of the fluctuational regime is of the order of $\delta r\approx0.1$
for realistic arrays. Note that the magnetic field that drives the
ladder across the QPT, $B<\Phi_{0}/A$, where $A$ is the area of
the unit cell, is very weak ($\sim1G$ in our experiments). 

We emphasize that Josephson ladders with $E_{JL}/E_{CL}\gg1$ have
exponentially small probability of phase slip processes in which the
phase across the ladder changes by $\sim2\pi$. This implies that
the static offset charges on individual islands have no effect on
the quantum states of the ladder. We shall neglect the effects of
these charges and phase slip processes in the following discussion.
The estimate for the phase slip rate can be found in \cite{Bell2012}. 

The low-energy physics of the ladders close to the QPT can be mapped
on the $\varphi^{4}$ model, which is relevant to a wide range of
physical phenomena, from quark confinement to ferromagnetism. Near
the critical point this model is equivalent to the integrable model
of 1D Ising spin chain in the transverse magnetic field. Flexibility
of the array design and tunability of the parameters of individual
Josephson junctions offer several unique opportunities for exploration
of the 1D quantum phase transitions. For example, this novel platform
facilitates the study of effects of non-integrability and disorder
on QPTs and the emergence of ergodic behavior in almost integrable
quantum systems.

\section{Theory of Long Josephon Ladders}

\subsection{Effective action of the low energy modes of the long chains \label{subsec:Effective-action-of}}

As we show below the long chain displays the phase transition that
occurs as a function of the ratio $r=E_{\textrm{JL}}/E_{\textrm{JS}}$
at a fixed flux through a unit cell, or at a fixed $r$ as a function
of the flux through each unit cell. As usual in phase transitions,
the critical properties are not sensitive to the parameter that drives
the transition, so we focus on the transition driven by $r$ below.
Small modifications introduced by transition driven by magentic field
were discussed in \cite{Bell2012}. The most important is the fact
that close to $r=r_{0}$ the effective distance from the critical
line is a quadratic function of $\Phi-\Phi_{0}/2$. 

\begin{figure}
\includegraphics[width=1\columnwidth]{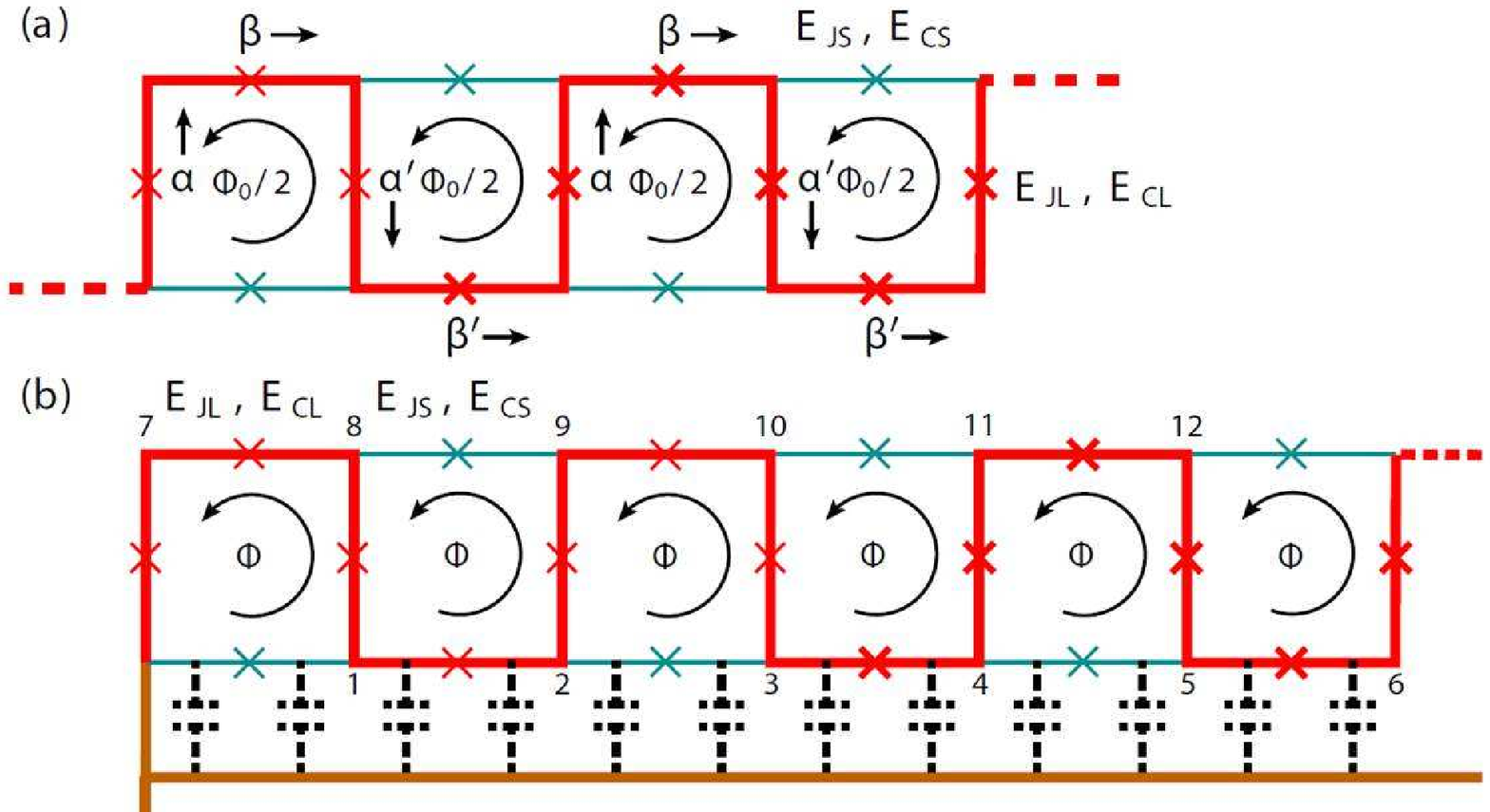}

\includegraphics[width=0.55\columnwidth]{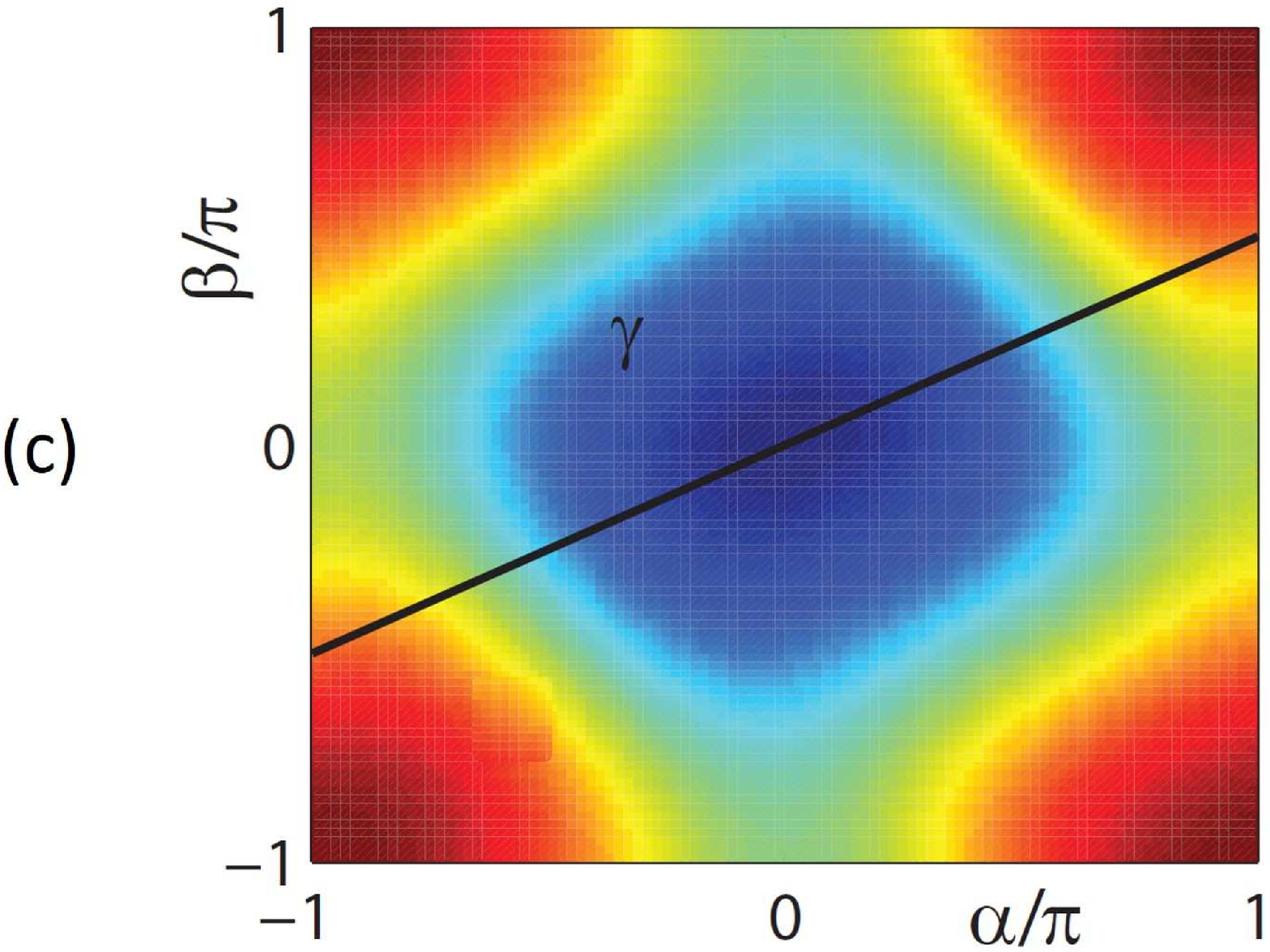}

\caption{Array schematics. Panel (a): exactly at full frustration the classical
ground state of the array corresponds to translationally invariant
pattern of phase differences $\alpha=\alpha',\beta=\beta'$ . Away
from the full frustation the period is doubled and phases $\alpha\protect\neq\alpha'$,
$\beta\protect\neq\beta'$. In the absence of the ground capacitance
the kinetic energy of the array depends only on local phase differences
$\alpha$ and $\beta$. For a generic fluctuation, the phases in different
unit cells become different, $\alpha_{j}\protect\neq\alpha$, etc.
Panel (b): the presence of a non-zero ground capacitance leads to
a small term that depends on the global phase $\Theta=\sum_{j<i}(\alpha_{j}+\beta_{j})$
that becomes relevant for long chains. Panel (c) shows the energy
as a function of the coordinates $\alpha,\beta$. Close to the minimum
the energy changes slowly in the direction indicated by the black
line.}
\label{schematics}
\end{figure}
We start by reviewing the properties of short chains in which one
can neglect the phase variations between adjacent elements and its
variation in time. Consider a small portion of the fully frustrated
ladder shown in Fig. \ref{schematics}. The translational invariance
implies that the solution is described by two phases across larger
junctions: $\alpha$ at the ``vertical'' junctions and $\beta$
at the ``horizontal'' ones. The energy per unit cell that contains
two larger and one smaller junctions is given by 
\begin{equation}
E(\alpha,\beta)=-E_{{\rm JL}}(\cos\alpha+\cos\beta)-E_{{\rm JS}}\cos(\pi-2\alpha-\beta).\label{eq:E(alpha,beta)}
\end{equation}

At $E_{\textrm{JL}}\gg E_{\textrm{JS}}$ the first term in Eq.~(\ref{eq:E(alpha,beta)})
dominates and $E(\alpha,\beta)$ has a minimum at $(\alpha,\beta)=(0,0)$.
The expansion near this point gives 
\begin{equation}
E^{(2)}(\alpha,\beta)=\frac{E_{JS}}{2}\,[\begin{array}{cc}
\alpha & \beta\end{array}]\left[\begin{array}{cc}
r-4 & -2\\
-2 & r-1
\end{array}\right]\left[\begin{array}{c}
\alpha\\
\beta
\end{array}\right],\label{eq: E^(2)(alpha,beta)}
\end{equation}
which is a quadratic form with eigenvalues $E_{{\rm JL}}$ and $(r-5)E_{{\rm JS}}$.
The second eigenvalue changes sign as the ratio $r=E_{\textrm{JL}}/E_{\textrm{JS}}$
decreases, signaling the instability and appearance of a nontrivial
ground state. The long chains display the properties similar to the
critical Ising model at $r$ slightly less than the critical value
of the ratio, $r_{\textrm{0}}=5$, as explained below\footnote{Short scale quantum fluctuations may renormalize the value of $r_{0}$
by shifting it down: $r_{0}=5-\delta r_{sh}$ This effect is numerically
small ($\delta r_{sh}\lesssim0.1$ ) for the chains studied experimentally
and qualitatively irrelevant so we neglect it in the following discussion. }. Exactly at $r_{0}=5$ the function $E(\alpha,\beta)$ becomes flat
along the eigenvector $(2,1)$ but remains steep in the orthogonal
direction (see Fig. \ref{schematics}c). We introduce new coordinates
in the ``flat'' and ``steep'' directions: $\gamma=(2\alpha+\beta)/\sqrt{5}$,
$\delta=(\alpha-2\beta)/\sqrt{5}.$ Figure~\ref{fig:elementary-cell}c
shows the energy plotted along the flat direction. Neglecting the
phase deviations in the steep direction, the total phase across the
unit cell is given by $\varphi_{\text{eb}}=3/\sqrt{5}\gamma$.

At $r>r_{\textrm{0}}$ the ground state values of phases $\gamma$
and $\varphi$, $\gamma_{0}$, and $\varphi_{0}$, are zero. Deviations
of the total phase from $\varphi_{0}=0$ result in the quadratic increase
of the energy 
\[
E^{(2)}(\varphi)=\frac{E_{\textrm{JL}}}{2}\delta r\gamma^{2}.
\]
where $\delta r=(r-r_{0})/r$. At the optimal point ($\delta r=0$)
the phase stiffness vanishes and the phase fluctuations are limited
by the next-order quartic term: 

\begin{equation}
E^{(4)}(\gamma)=\frac{1}{24}\left(25-\frac{17r}{25}\right)E_{\textrm{JS}}\gamma^{4}\approx0.9E_{\textrm{JS}}\gamma^{4}\label{eq:E^4}
\end{equation}
where in the last equality we replaced the numerical coefficient by
it value at $r=r_{0}$. At smaller $r<r_{\textrm{0}}$ the ground
state corresponds to a non-zero value of $|\gamma_{0}(r)|\propto\sqrt{\delta r}$
. The appearance of a non-zero $\gamma_{0}$ implies a phase transition
that breaks $Z_{2}$ symmetry $\gamma\rightarrow-\gamma$. 

Close to the phase transition the phase fluctuations become relevant.
We derive the action of these fluctuations assuming that they are
slow in space and time. We begin with spatial fluctuations. For a
general space varying phase one gets
\begin{align*}
E & =-E_{{\rm JL}}\sum_{i}(\cos\alpha_{i-1/2}+\cos\alpha_{i+1/2}+\cos\beta_{i})\\
-E_{{\rm JS}} & \sum_{i}\cos(\pi-\alpha_{i-1/2}-\alpha_{i+1/2}-\beta_{i})
\end{align*}
where we label phases of each wire by the coordinates of their centers,
so that the phases of horizontal wires have integer coodinates and
the centers of the vertical ones have half-integer coordinates. Expanding
in the phases $\alpha_{i},\beta_{i}\ll1$ and taking the Fourrier
transform we get the quadratic part of the energy of the fluctuation
with momentum $k$:
\begin{equation}
E^{(2)}(k)=\frac{E_{\textrm{JS}}}{2}\sum_{k}\left[\alpha_{k},\beta_{k}\right]\left[\begin{array}{cc}
r+2\xi_{k} & \xi_{k}\\
\xi_{k} & r-1
\end{array}\right]\left[\begin{array}{c}
\alpha_{-k}\\
\beta_{-k}
\end{array}\right].\label{eq:E^(2)(k)}
\end{equation}
where $\xi_{k}=-2\cos(k/2)$. Expanding it for small $k$ and leaving
only the contribution of the slow mode $\alpha=2/\sqrt{5}\gamma$
and $\beta=1/\sqrt{5}\gamma$ we get for the energy per unit cell
\begin{equation}
E^{(2)}(k)=\frac{1}{2}E_{\textrm{JL}}\left[\frac{3}{5r}(\nabla\gamma)^{2}+\delta r\gamma^{2}\right].\label{eq:E^2(k)}
\end{equation}
The kinetic energy per unit cell is due to the capacitances of the
larger and smaller junctions. We assume that $C_{L}=rC_{S}$ and express
the result in the units of $E_{CL}=(2e)^{2}/C_{L}$: 
\begin{align}
T & =\frac{1}{2E_{CL}}\left[\left(\frac{d\alpha}{dt}\right)^{2}+\left(\frac{d\beta}{dt}\right)^{2}+\frac{1}{r}\left(\frac{d\left(2\alpha+\beta\right)}{dt}\right)^{2}\right]\nonumber \\
 & =\frac{(r+5)}{2rE_{CL}}\left(\frac{d\gamma}{dt}\right)^{2}\label{eq:T}
\end{align}

Away from the critical point the interaction between fluctuations
can be neglected. In this regime the spectrum of the slow fluctuations
characterized by (\ref{eq:E^2(k)},\ref{eq:T}) is given by the relativistic
dependence
\begin{align}
\omega^{2} & =\left(ck\right)^{2}+m^{2}\label{eq:SpectrumLinearApp}\\
c^{2} & =\frac{3}{5(r+5)}E_{JL}E_{CL}\nonumber \\
m^{2} & =\frac{r\delta r}{5(r+5)}E_{JL}E_{CL}.\nonumber 
\end{align}

Note that at all $r$ the velocity is relatively small because it
is due to the Josephson energy of the smaller junctions and the charging
energy of the larger ones, so that the wave dispersion is always small
in the Josephson chain of this type. As the optimal point $\delta r=0$
is approached, the gap in the spectrum closes and the fluctuations
become more relevant. At large $E_{JL}/E_{CL}\gg1$ the width of the
fluctuational regime, $\delta r^{*}$, where the effects of the interaction
(\ref{eq:E^4}) are relevant, is small. In the following discussion
of this regime we assume $r\approx r_{0}=5$ so that the full action
becomes
\begin{align*}
S & =\int\mathcal{L}dtdx\\
\mathcal{L} & =\frac{1}{E_{CL}}\left(\frac{d\gamma}{dt}\right)^{2}+\frac{E_{\textrm{JL}}}{2}\left[\frac{3}{25}(\nabla\gamma)^{2}+\delta r\gamma^{2}+\frac{9}{25}\gamma^{4}\right]
\end{align*}
 Rescaling the variables 
\begin{align*}
x & =\frac{\sqrt{3}}{5}s\\
t & =\sqrt{\frac{2}{E_{JL}E_{CL}}}\tau\\
\gamma & =\frac{5}{3\sqrt{2}}\eta
\end{align*}
we reduce it to a fully dimensionless form
\begin{equation}
S=\frac{1}{2g}\int\left\{ \left(\frac{d\eta}{d\tau}\right)^{2}+(\nabla\eta)^{2}+\delta r\eta^{2}+\frac{1}{2}\eta^{4}\right\} d\tau dx\label{eq:S}
\end{equation}
where 
\begin{equation}
g=\frac{3\sqrt{6}}{5}\sqrt{\frac{E_{CL}}{E_{JL}}}.\label{eq:g}
\end{equation}

We estimate the width of the fluctuational regime where the spectrum
(\ref{eq:SpectrumLinearApp}) is not valid by computing the fluctuational
corrections to the action (\ref{eq:S}). The leading correction to
the coefficient of $\eta^{4}$ term is 
\[
\frac{9g}{8\pi\delta r},
\]
so that the width of the fluctuational regime $\delta r^{*}\approx g.$ 

\subsection{Mapping to the Ising model\label{subsec:Mapping-to-Ising}}

The transition to the ordered state breaks $Z_{2}$ symmetry and belongs
to the same universality class as the Ising model. In the ordered
state the field $\eta$ acquires average $\eta=\pm\eta_{0}$ where
$\eta_{0}=(-\delta r)^{1/2}$; in this regime the low energy excitations
are domain walls with the static energy 
\[
\epsilon_{0}=\frac{1}{\sqrt{2}}\frac{1}{g}\delta r^{3/2}.
\]

Because the time needed to create and destroy a domain wall is $\tau=\delta r^{-1/2},$
the action correponding to this process is $S\sim\delta r/g$ which
again tells us that the width of the fluctuational regime is $\delta r^{*}\approx g$. 

Outside of the fluctuational regime the kinetic energy of the domain
wall moving with velocity $v$ is 
\[
\frac{v^{2}}{2\sqrt{2}g}\delta r^{3/2}
\]
implying the mass $m=\delta r^{3/2}/\sqrt{2}g$. The Lorentz invariant
equation compatible with these limits for the domain wall energy spectrum
is 
\[
\epsilon(k)=(m^{2}+k^{2})^{1/2}.
\]
In the ordered phase and inside the critical fluctuational regime
the model should be similar to the quantum Ising model 
\[
H=\mathfrak{\mathscr{T}}\sum_{i}\sigma_{i}^{x}+J\sum_{i}\sigma_{i}^{z}\sigma_{i}^{z}
\]
 where $t$ and $J$ are smooth functions of $\delta r$. Comparing
the spectra of the domain wall in the Ising model and action (\ref{eq:S})
we see that 
\begin{align*}
J & \approx m/2\\
\mathfrak{\mathscr{T}} & \approx\frac{1}{2ma^{2}}
\end{align*}
in the ordered phase at $J\gg\mathfrak{\mathscr{T}}$ where $a$ is
the distance between the Ising spins. The distance $a$ is set by
the domain wall size $a^{2}=1/\left|\delta r\right|$ . 

The Ising model has transition to the ordered state at $J=\mathfrak{\mathscr{T}}$.
By translating it into the parameters of the original model we see
that the transition happens at 
\[
\delta r_{c}=-cg
\]
 where $c\sim1$. 

Close to the transition into the ordered state the excitations of
the Ising model are Majorana fermions 
\begin{align*}
c_{p}^{\dagger}(k) & =(-1)^{\sum_{j<k}n_{p}\left(j\right)}\sigma_{p}^{+}\left(k\right)\\
n_{p}(j) & =\left[\sigma_{j}^{z}(p)+1\right]/2.
\end{align*}

Qualitatively these excitations describe the domain walls in the Ising
model that propagate with a constant velocity exactly at the transition,
i.e. they have relativistic spectrum. The fermionic nature of these
excitations describes the exclusion principle of domain walls. Generally,
away from the transition point the spectrum of these excitations is
the same as for free but massive one dimensional fermions:
\begin{equation}
\epsilon(k)=2\sqrt{\mathfrak{\mathscr{T}}^{2}+J^{2}-2\mathfrak{\mathscr{T}}J\cos(ka)}.\label{eq:epsilon(k)}
\end{equation}

\begin{figure}
\includegraphics[width=0.8\columnwidth]{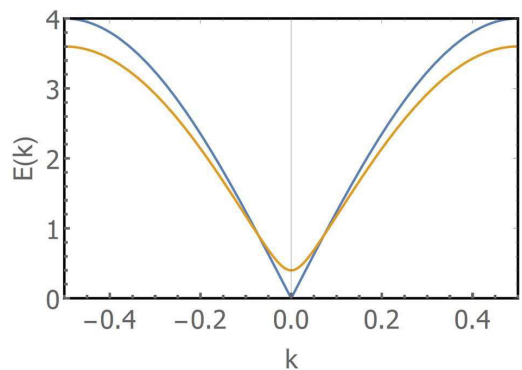}

\caption{Characteristic spectra of the excitations in the Ising model at and
slightly away from the transition (blue and yellow curves, respectively).
At low energies the spectra become relativistic with the mass that
vanishes at the transition. In the presence of a non-zero ground capacitance
the spectra as a function of $k$ remain qualitatively similar. }

\end{figure}
We now discuss mapping of the matrix elements of the physical operators
to the operators of the Ising model. The physical currents through
the individual junctions are $J=E_{JL}\sin\alpha$, $J=E_{JL}\sin\beta$
and $J=-E_{JS}\sin(2\alpha+\beta)$. Thus, the current operator is
generally given by $J=c_{J}E_{JL}\eta$ where $c_{J}\sim1$ is a numerical
coefficient that depends on the particular current studied. In the
critical regime in which locally $\eta=\pm\eta_{0}$ so the current
operator is directly related to the Ising variable 
\begin{equation}
\hat{J}_{eff}=c_{J}E_{JL}\eta_{0}\sigma^{z}.\label{eq:J_eff}
\end{equation}
 External flux noises interact directly with the current operators
by $H_{noise}=\delta\Phi_{ext}\hat{J}$. Below we discuss the effect
of this interaction on the decay and dephasing of Majorana modes with
spectrum (\ref{eq:epsilon(k)}). 

The charge operator associated with larger and smaller junctions are
$Q_{L1}=id/d\alpha,$ $Q_{L2}=id/d\beta,$ $Q_{S}=id/d(2\alpha+\beta).$
These charges interact with the potentials created by stray charges
on each island. As before, we focus on the effect of this interaction
with the soft modes described by the Ising variables. Consider a single
unit cell described by the local Hamiltonian 
\[
H=-\frac{E_{CL}}{4}\left(\frac{d}{d\gamma}\right)^{2}+\frac{1}{2}E_{\textrm{JL}}\left[+\delta r\gamma^{2}+\frac{9}{25}\gamma^{4}\right]
\]
This Hamiltonian has two low energy states $\Psi_{\pm}(\gamma)$ that
correspond to symmetric and antisymmetric states that become eigenstates
of Ising operator $\sigma_{x}$ after mapping to the Ising model.
These states are split by the energy $\delta E=E_{-}-E_{+}$ that
becomes $\delta E=2t$ in the Ising model. Assuming that $\delta E\ll\sqrt{E_{CL}E_{L}}$,
i.e. that the Ising modes are well separated from the rest of the
spectrum, we can evaluate the matrix element of the charge operator
$Q_{\gamma}=id/d\gamma$ between these states:
\[
\left\langle +\left|Q_{\gamma}\right|-\right\rangle =i\frac{\mathfrak{\mathscr{T}}}{E_{c}}\eta_{0}
\]

The symmetry of the states guarantees that all other matrix elements
of the charge are zero, thus 
\begin{equation}
\hat{Q}_{eff}=\frac{\mathfrak{\mathscr{T}}}{E_{c}}\eta_{0}\sigma^{y}.\label{eq:Q_eff}
\end{equation}

Note that the matrix element of the charge operator contains two factors
that are small in the regime where the model (\ref{eq:S}) can be
mapped to Ising model: $\eta_{0}\ll1$ and $\mathfrak{\mathscr{T}}/E_{c}\ll1$. 

\subsection{Matrix elements of the physical operators close to the critical point\label{subsec:Matrix-elements-of} }

Close to the critical point the matrix elements of both the external
charge and flux are given by operators that act on the effective Ising
variables (\ref{eq:J_eff},\ref{eq:Q_eff}). We evaluate the effect
of these operators on the low energy (Majorana) modes of the critical
Ising model. To compute the matrix elements of the $\sigma_{i}^{y}$
operator between the ground-state and low-energy excited states, it
is convenient to use the Jordan-Wigner transformation in the form:
\begin{eqnarray}
\sigma_{i}^{x} & = & 1-2c_{i}^{\dagger}c_{i}\\
\sigma_{i}^{z} & = & \left(\prod_{j=1}^{i-1}(1-2c_{j}^{\dagger}c_{j})\right)(c_{i}+c_{i}^{\dagger})\\
\sigma_{i}^{y} & = & \left(\prod_{j=1}^{i-1}(1-2c_{j}^{\dagger}c_{j})\right)i(c_{i}-c_{i}^{\dagger})
\end{eqnarray}
Here it is assumed that we have a ladder with $N$ unit cells, so
the index $i$ runs from 1 to $N$. Note that the Ising Hamiltonian
is changed into its opposite after performing a $\pi$ rotation around
the $x$ axis of the spins on one sublattice (e.g. for $i$ even).
Doing this changes the dispersion relation into: 
\begin{equation}
\epsilon(k)=2\sqrt{\mathfrak{\mathfrak{\mathscr{T}}}^{2}+J^{2}-2\mathscr{T}J\cos(ka)}
\end{equation}
which is convenient, because the minimum of this dispersion relation
occurs at $k=0$. Introducing $A_{i}=c_{i}+c_{i}^{\dagger}$ and $B_{i}=c_{i}^{\dagger}-c_{i}$
allows then to write the Ising Hamiltonian as: 
\begin{equation}
H=-\mathfrak{\mathscr{\mathscr{T}}}\sum_{i=1}^{N}A_{i}B_{i}-J\sum_{i=1}^{N-1}B_{i}A_{i+1}.
\end{equation}
The single-excitation creation and annihilation operators $\gamma_{k}^{\dagger}$,
$\gamma_{k}$ are written as: 
\begin{multline}
\gamma_{k}=\sum_{j=1}^{N}(\phi_{j}^{k}A_{j}+\chi_{j}^{k}B_{j})\\
\gamma_{k}^{\dagger}=\sum_{j=1}^{N}(\phi_{j}^{k}A_{j}-\chi_{j}^{k}B_{j})\label{eq:gamma_k}
\end{multline}
where the coefficients $\phi_{j}^{k}$ and $\chi_{j}^{k}$ are real.
These operators are determined by the condition that $[H,\gamma_{k}]=-\epsilon(k)\gamma_{k}$,
which gives: 
\begin{eqnarray}
2t\phi_{j}^{k}-2J\phi_{j+1}^{k} & = & -\epsilon(k)\chi_{j}^{k}\\
2t\chi_{j}^{k}-2J\chi_{j-1}^{k} & = & -\epsilon(k)\phi_{j}^{k}
\end{eqnarray}
With the open boundary conditions, we have to solve these equations
together with the constraints $\chi_{0}^{k}=0$ and $\phi_{N+1}^{k}=0$.
The allowed values of $k$ are determined by: 
\begin{equation}
\frac{\sin(kaN)}{\sin(ka(N+1))}=\frac{\mathfrak{\mathscr{T}}}{J}
\end{equation}
Note that $k=0$ and $k=\pi$ are both excluded, because they lead
to vanishing amplitudes $\phi_{j}^{k}$ and $\chi_{j}^{k}$. Therefore,
we have to search for real solutions in the interval $0<k<\pi$. There
are exactly $N$ such solutions when $\mathfrak{\mathscr{T}}/J>\frac{N}{N+1}$,
which corresponds to the paramagnetic phase. In the other case $\mathfrak{\mathscr{T}}/J<\frac{N}{N+1}$,
we get only $N-1$ real solutions and one complex solution $k=i\kappa$,
where $\kappa$ is given by: 
\begin{equation}
\frac{\sinh(kaN)}{\sinh(ka(N+1))}=\frac{\mathfrak{\mathscr{T}}}{J}
\end{equation}
and the corresponding energy is: 
\begin{equation}
\epsilon(i\kappa)=2\sqrt{\mathfrak{\mathscr{T}}^{2}+J^{2}-2\mathfrak{\mathscr{T}}J\cosh(\kappa a)}
\end{equation}
which lies inside the gap of the infinite chain. The real solutions
have the form: 
\begin{eqnarray}
\phi_{j}^{k} & = & \phi_{0}^{k}\frac{\sin(ka(N+1-j))}{\sin(ka(N+1))}\\
\chi_{j}^{k} & = & -\phi_{0}^{k}\frac{\epsilon(k)}{2J}\frac{\sin(kaj)}{\sin(ka)}
\end{eqnarray}
and the bound state wave-function is obtained by analytical continuation
from $k$ to $i\kappa$. Canonical fermionic anticommutation relations
are satisfied provided: 
\begin{equation}
\sum_{j=1}^{N}\phi_{j}^{k}\phi_{j}^{k'}=\sum_{j=1}^{N}\chi_{j}^{k}\chi_{j}^{k'}=\frac{\delta_{kk'}}{4}
\end{equation}
When $k=k'$, these relations fix the value of $\phi_{0}^{k}$ in
the above expressions for $\phi_{j}^{k}$ and $\chi_{j}^{k}$. Completeness
can then be used to show that: 
\begin{eqnarray}
A_{j} & = & 2\sum_{k}\phi_{j}^{k}(\gamma_{k}+\gamma_{k}^{\dagger})\label{expr_A_j}\\
B_{j} & = & 2\sum_{k}\chi_{j}^{k}(\gamma_{k}-\gamma_{k}^{\dagger})\label{expr_B_j}
\end{eqnarray}
Note that the $k$ sums have to incorporate the bound-state when $\mathfrak{\mathscr{T}}/J<\frac{N}{N+1}$.

\begin{figure*}
\includegraphics[width=1\textwidth]{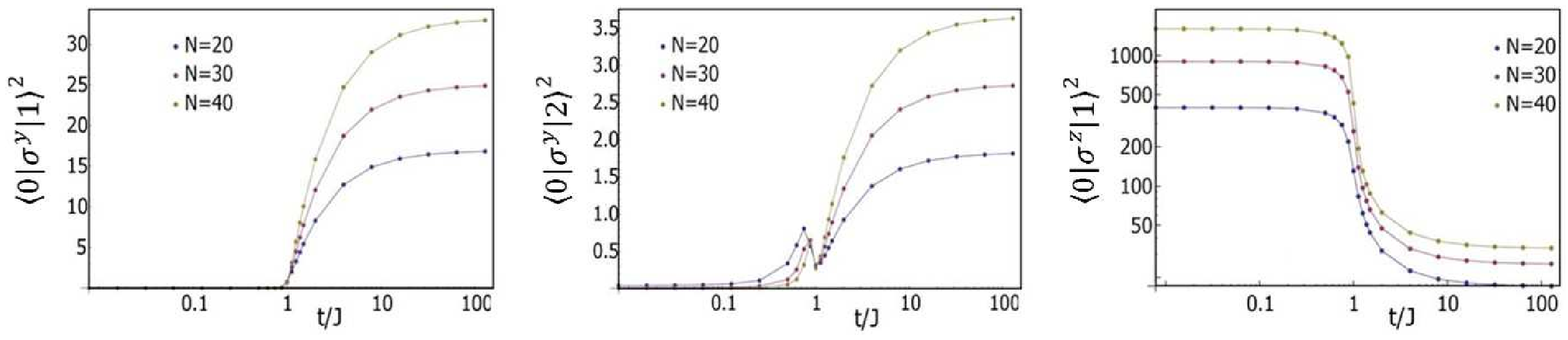}

\caption{Matrix elements of the relevant physical operators between the lowest
energy states close to quantum critical point of the Ising model,
from left to right: the matrix element of the dimensionless charge
operator $\sigma^{y}$ between two lowest states, the same matrix
element between the ground and the second excited state, and the matrix
element of flux noise $\sigma^{z}$ between the ground state and the
lowest excited state. \label{fig:Matrix-elements-of}}
\end{figure*}

The ground state of the system $\left|0\right\rangle $ is determined
by the condition that $\gamma_{k}\left|0\right\rangle =0$ for all
allowed $k$'s. As shown earlier, we have to evaluate the matrix element
of $\sigma_{i}^{y}$ between the ground state and low lying states.
The Jordan-Wigner representation for $\sigma_{i}^{y}$ can also be
written as: 
\begin{equation}
\sigma_{i}^{y}=-i\left(\prod_{j=1}^{i-1}A_{j}B_{j}\right)B_{i}
\end{equation}
so it is the product of an odd number of fermionic operators. Therefore,
$\sigma_{i}^{y}$ couples the ground state to excited states containing
an odd number of elementary fermionic excitations. The simplest of
these states are of the form $\gamma_{k}^{\dagger}\left|0\right\rangle $.
Evaluation of the matrix elements is straighforward, using Wick's
theorem. For this, we need the basic correlators, which are easily
determined from~(\ref{expr_A_j}),~(\ref{expr_B_j}) and the completeness
relations: 
\begin{eqnarray}
\left\langle 0|A_{i}A_{j}|0\right\rangle  & = & \delta_{ij}\\
\left\langle 0|B_{i}B_{j}|0\right\rangle  & = & -\delta_{ij}\\
\left\langle 0|A_{i}B_{j}|0\right\rangle  & = & -4\sum_{k}\phi_{i}^{k}\chi_{j}^{k}\label{G_ij}\\
\langle0|B_{i}\gamma_{k}^{\dagger}|0\rangle & = & 2\chi_{j}^{k}\label{D_ik}
\end{eqnarray}
Because the correlations of $A_{j}$ operators with themselves are
purely local, and similarly for the correlations of $B_{j}$ operators,
we see that upon computing $\langle0|\sigma_{i}^{y}\gamma_{k}^{\dagger}|0\rangle$,
each $A_{i}$ has to be paired with a $B_{j}$, and $\gamma_{k}^{\dagger}$
has to be paired with the remaining $B_{j}$ operator. This leads
to the determinantal formula: 
\begin{equation}
\langle0|\sigma_{i}^{y}\gamma_{k}^{\dagger}|0\rangle=-i\mathrm{Det}C_{i}^{k}
\end{equation}
Here, $C_{i}^{k}$ is an $i\times i$ matrix given by: 
\begin{equation}
C_{i}^{k}=\left(\begin{array}{ccccc}
G_{1,1} & G_{1,2} & G_{1,3} & ... & G_{1,i}\\
G_{2,1} & G_{2,2} & G_{2,3} & ... & G_{2,i}\\
... & ... & ... & ... & ...\\
G_{i-1,1} & G_{i-1,2} & G_{i-1,3} & ... & G_{i-1,i}\\
D_{1}^{k} & D_{2}^{k} & D_{3}^{k} & ... & D_{i}^{k}
\end{array}\right)
\end{equation}
where $G_{i,j}=\langle0|A_{i}B_{j}|0\rangle$ and $D_{i}^{k}=\langle0|B_{i}\gamma_{k}^{\dagger}|0\rangle$,
which are given by Eqs.~(\ref{G_ij}) and~(\ref{D_ik}), respectively.
Such determinantal expressions for the Ising form factors are well-known,
see e.g. \cite{Konik1996}. For their actual evaluation on a finite
size chain with open boundary conditions, we used numerics. The results
are shown in Fig \ref{fig:Matrix-elements-of}. As one might expect,
the flux operator proportional to $\sigma^{z}$ acquires large matrix
elements in the ordered state. Because the ground state in a finite
system is a symmetric combination of up and down ferromagnetically
ordered state, the non-zero matrix elements correspond to the 01 transitions.
Similarly, the charge operator acquires large matrix elements between
ground and the first excited states in the disordered state of the
Ising model. In both cases these matrix elements imply the decay rate
of the first excited state. The matrix elements of the higher energy
states display similar behavior. 

\subsection{Effect of a non-zero ground capacitance}

In a realistic ladder of Josephson junctions each element of the ladder
has a non-zero ground capacitance as shown schematically in Fig.\ref{schematics}b.
For instance, in the ladders implemented experimentally, the ratio
$C_{G}/C_{L}\approx0.05-0.1$. Though small, such capacitance has
significant effect on the low-energy spectrum because it introduces
the terms in the effective action of the unit cell that depend on
the time derivative of the total phase:
\begin{equation}
\delta L=\frac{1}{2E_{CG}}\left(\frac{d\Theta}{dt}\right)^{2}\label{eq:deltaL}
\end{equation}
 where $E_{CG}=(2e)^{2}/C_{G}$ and the total phase $\Theta$ is related
to the low energy mode variable $\gamma$ by $\triangledown\Theta=3/\sqrt{5}\gamma$. 

The full kinetic energy of the low energy modes becomes
\begin{equation}
T=\frac{(r+5)}{2rE_{CL}}\left[1+\frac{\kappa^{2}}{k^{2}}\right]\left(\frac{d\gamma}{dt}\right)^{2}\label{eq:T_Cg}
\end{equation}
where 
\begin{equation}
\kappa^{2}=\frac{9r}{5(r+5)}\frac{E_{CL}}{E_{CG}}\ll1\label{eq:kappa}
\end{equation}
is the inverse characteristic length associated with the ground capacitance. 

The appearance of the length $\kappa^{-1}$ (\ref{eq:kappa}) implies
that the effective Coulomb interaction between the charges is screened
at distances $L>\kappa^{-1}$, in terms of the charge variable, $q_{\gamma}$
conjugated to $\gamma$ the energy (\ref{eq:T_Cg}) becomes
\begin{equation}
T=\frac{rE_{CL}}{2(r+5)}\frac{k^{2}}{k^{2}+\kappa^{2}}q_{\gamma}^{2}.\label{eq:T_q}
\end{equation}

In the linear regime where non-linear effects can be neglected, this
addition changes the spectrum (\ref{eq:SpectrumLinearApp}) to become
\begin{equation}
\omega^{2}=\frac{(ck)^{2}+m^{2}}{1+\frac{\kappa^{2}}{k^{2}}}\label{eq:SpectrumWithGC}
\end{equation}
 that displays additional softnening at low wave vectors $ck\ll(c\kappa,m)$:
$\omega\approx m\kappa^{-1}k$. The softening of the spectrum (\ref{eq:SpectrumWithGC})
is observable only if it occurs outside of the critical regime, i.e.
at $\delta r>g$ that corresponds to $\kappa>g\approx(E_{CL}/E_{JL})^{1/2}$. 

In order to evaluate the effect of the ground capacitance on the low-energy
modes in the critical regime, we evaluate the effect of the additional
kinetic energy (\ref{eq:deltaL}) on the effective action of the Majorana
quasiparticles. It is convenient to represent the action as due to
the interaction with the scalar field $\phi$: 
\begin{equation}
\delta L=i\frac{d\eta}{dt}\phi+\frac{E_{G}}{2}\left[(\triangledown\phi)^{2}+\kappa^{2}\phi^{2}\right]\label{eq:deltaL_decoupled}
\end{equation}
where $E_{G}=(2/5)E_{CG}$. The last term in (\ref{eq:deltaL_decoupled})
is due to the screening of the Coulomb interaction at short scales
discussed above. By repeating the same arguments that led us to (\ref{eq:Q_eff})
for the charge operator, the additional terms in Lagrangian density
(\ref{eq:deltaL_decoupled}) can be translated into
\begin{equation}
\delta H=\mathfrak{\mathscr{T}}\eta_{0}\sigma_{i}^{x}\phi(x_{i})+\frac{E_{G}}{2}\left[(\triangledown\phi)^{2}+\kappa^{2}\phi^{2}\right]\label{eq:delta_H}
\end{equation}
for the effective Ising model that describes the vicinity of the transition
point. 

In the fermionic representation these terms imply weak residual interaction
between fermionic densities $\sigma_{i}^{x}=1-2c_{i}^{\dagger}c_{i}$.
This interaction leads to the self-energy correction, $\Sigma(k)$,
to the fermionic Green function. The important property of the long
range interaction between spins or fermionic modes (\ref{eq:gamma_k})
is its dependence on the momentum $k$ but not $\omega$. Physically
it is due to the instanteneous and long-range nature of Coulomb interaction.
Such interaction leads to the self-energy corrections that depend
on the momentum but not on the frequency and thus violate the Lorentz
invariance of the low-energy spectra. Qualitatively, the effect of
the Coulomb interaction is to attract the spin flips (fermion densities)
at short scales. The appearance of the frequency-independent $\Sigma(k)$
implies that the minimum of the energy $\omega_{k}(r)$ occurs at
different $r$ for different $k$. 

In the experiment discussed in Section \ref{sec:Experimental-data}
we study the spectrum dependence on the flux, not on $r.$ This is
equivalent to the motion in the horizontal direction in the phase
diagram of Fig. \ref{fig:elementary-cell}d instead of the vertical
one. The shift in $r$ of the position of the minima implies that
the corresponding parabolas at which $\omega_{k}(r,\Phi)$ has a minimum
are shifted up for larger $k$ that translates into the downshift
in the value of the flux at which $\omega_{k}(\Phi)$ has a minimum
for constant $r$. 

The analytical expressions for the shift of $\omega_{k}(r,\Phi)$
can be obtained close to the Ising critical point. Exactly at the
critical point of the Ising model without corrections (\ref{eq:delta_H})
the fermionic density has power law correlators that translates into
power law correlations 
\[
\left\langle \sigma_{0}^{x}\sigma_{r}^{x}\right\rangle =\frac{1}{\pi^{2}r^{2}}
\]
 which can be described as the correlator of the free bosonic fields,
$s$. In this hydrodynamic description the addition of the terms (\ref{eq:delta_H})
does not change the quadratic nature of the energy. The Green function
$D(\omega,k)$ of the bose field acquires a self energy 
\[
\Sigma(k)=\frac{\left(\mathfrak{\mathscr{T}}\eta_{0}\right)^{2}}{E_{G}}\frac{1}{k^{2}+\kappa^{2}}
\]
 that has no frequency dependence. This self energy correction has
exactly the effect discussed above: it shifts the position of singularity
in $D(\omega,k$) away from $\omega=\epsilon(k)$ line that implies
the shift in the spectrum of elementary (fermionic) excitations. In
particular, the positions of the minuma, $r_{min},$ of $\omega_{k}(r,\Phi)$
as a function of $r$ for different $k$ occur at different $r$.
The effect is maximal for $k\sim\kappa$ at which $\Sigma\sim cg$
that leads to the shift in $r_{min}\sim g$. 

\section{Experimental data\label{sec:Experimental-data}}

\begin{figure}
\includegraphics[width=1\columnwidth]{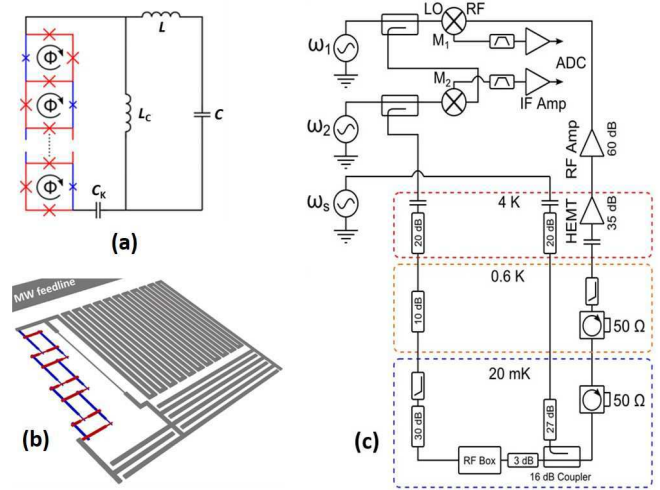}

\caption{Simplified diagram of the microwave setup \cite{Bell2012}. Panel
(a): Schematic description of the on-chip circuit. The Josephson ladder,
in combination with the capacitor $C_{K}$, forms a resonator which
is coupled to the readout $LC$ resonator via the kinetic inductance
$L_{C}$ of a narrow superconducting wire. Panel (b): The on-chip
circuit layout of the tested circuit inductively coupled to the microwave
(MW) feedline. Panel (c): Simplified circuit diagram of the measurement
setup. }

\label{MW setup}
\end{figure}
\begin{figure}
\includegraphics[width=1\columnwidth]{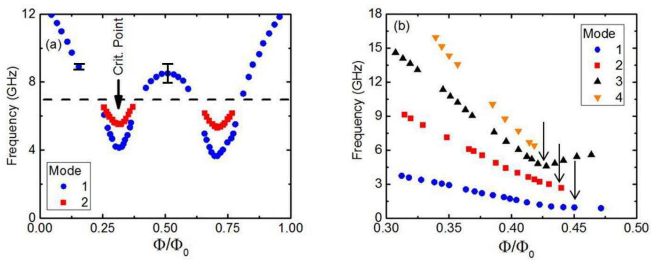}

\caption{Two-tone spectroscopy of the low-energy internal modes of the ladders
with $r\le r_{0}$ as a function of the magnetic flux in the unit
cells. Panel (a): The data for a 24-cell ladder with $r=3.2$. The
resonance modes are periodic in the flux with the period $\Phi_{0}$.
Two critical points are located symmetrically with respect to $\Phi=\Phi_{0}/2$.
At the critical points the single-minimum dependence $E_{J}(\varphi)$
is transformed into a double-minima function; this corresponds to
the QPT between the \textquotedblleft paramagnetic\textquotedblright{}
and \textquotedblleft ferromagnetic\textquotedblright{} phases. Note
the difference in the width of the resonances in the ``paramagnetic''
and ``ferromagnetic'' phases (shown as the error bars for two data
points). Panel (b): The data for a 92-cell ladder with $r=3.2$. The
resonance frequencies of several low-energy modes are shown near the
critical point. With approaching the critical point at $\Phi\approx0.45\Phi_{0}$
we observed pronounced softening of the low-energy modes of the ladder.
The resonances in the ferromagnetic phase are very broad (see the
text), so the accuracy of finding the critical points is low. The
minima positions can be extracted reliably from the remaining points
by fitting with a polynomial curve. Arrows indicate the critical points,
$\Phi_{C}(k)$, that are shifted for different modes due to the long-range
interactions between the array\textquoteright s unit cells. In the
limit of an infinitely long chain we expect the critical point to
be located at $\Phi=0.46\Phi_{0}$.}

\label{MW data on chain spectra}
\end{figure}
Our microwave measurements were designed to explore the spectrum of
low-energy modes of Josephson ladders near the 1D QPT and the ``lifetime''
of these modes. The experimental set-up for these spectral and time-domain
measurements has been described in Refs. \cite{Bell2014,Bell2016}.
Briefly, in these experiments the Josephson ladder was coupled to
the lumped-element readout resonator via the kinetic inductance $L_{C}$
of a narrow superconducting film (Figs. \ref{MW setup}a,b). The ladders,
the readout circuits, and the microwave (MW) transmission line on
the chip were fabricated using multi-angle electron-beam deposition
of Al films through a lift-off mask (for fabrication details see Refs.
\cite{Bell2012}). The in-plane dimensions of the Josephson junctions
varied between $0.1\times0.1\mu m^{2}$ and $0.3\times0.3\mu m^{2}$;
the area of a unit cell was $15\mu m^{2}$. The global magnetic field,
which determines the fluxes in all superconducting loops, has been
generated by a superconducting solenoid. 

In the dispersive two-tone measurements of the ladder spectra, the
resonance of the readout resonator was monitored at the probe frequency
$f_{1}$ while the ladder was excited by the second-tone (``pump'')
frequency $f_{S}$ (Fig. \ref{MW setup}c). The microwaves were transmitted
through the microstrip line coupled to the Josephson ladder and the
$LC$ resonator. The amplified signal was mixed by mixer M1 with the
local oscillator signal at frequency $f_{2}$. The intermediate-frequency
signal at $\Omega=f_{1}-f_{2}=30MHz$ was digitized by a 1 GS/s digitizing
card. The signal was digitally multiplied by $\sin(\Omega t)$ and
$\cos(\Omega t)$, averaged over an integer number of periods, and
its amplitude $A$ and phase were extracted. The reference phase (which
randomly changes when both $f_{1}$ and $f_{2}$ are varied in measurements)
was found using similar processing of the low-noise signal provided
by mixer M2 and digitized by the second channel of the ADC. Excitation
of the modes resulted in a change of the ladder impedance \cite{Wallraff2005};
this change was registered as a shift of the resonance of the readout
resonator probed at $f_{1}$. The low-energy modes were measured as
a function of the flux in the ladder's unit cells within the frequency
range of our microwave setup $0.5-20GHz$. 

In the time-domain experiments, we have observed Rabi oscillations
for a \textquotedblleft qubit\textquotedblright{} formed by the ladder
and a shunting capacitance $C_{K}$. The ladder modes were excited
by a short ($\sim0.02-10\mu s$) MW pulse at the second-tone frequency
$f_{S}$ and the population of the excited level was recorded at the
end of each pulse; the data were averaged over many repetitive measurements.
The Rabi time for the lowest-energy mode exceeded $2\mu s$ for the
arrays with 92 unit cells. This observation demonstrates that the
multi-junction arrays can be considered as quantum (not limited by
decoherence) systems over a relatively long time scale. These measurements
will be discussed elsewhere, below we focus on the spectroscopic data.

The results of measurements of the resonance frequencies for several
low-energy modes of the ladder with $r<r_{0}$ are shown in Fig. \ref{MW data on chain spectra}.
The nominal parameters \footnote{The junction energies given below were computed by using the data
for the junction area, measured resistance of the test junctions and
Ambegaokar-Baratoff relation. This computation gives reliable values
for $E_{JL}$ and $E_{CL}$ but significantly overestimates the value
of the Josephson energy for smaller junctions. } of the junctions in this chain were
\begin{multline}
E_{JS}=4.8\text{ K, }E_{CS}=0.41\text{ K}\\
E_{JL}=15.6\text{ K, }E_{CL}=0.14\text{ K}\label{eq:JJParameters}
\end{multline}
For these parameters we estimate $g\approx0.1$. The ground capacitance
in this experiment was $C_{G}\sim0.05C_{L}$, so the characteristic
length due to Coulomb interactions is $\kappa^{-1}\approx5\times(unit\,cell\,length)$.
For these parameters one expects to observe the relativistic spectrum
of the low-energy modes with $c\sim(10-20)[GHz]\times a_{0}$ where
$a_{0}\sim3\times(unit\,cell\,length)$; this would lead to the frequency
difference between the lowest modes $\Delta f=1-2\,\text{GHz}$ in
good agreement with our observations. The frequency shift (defined
as the difference between the minimal frequency of a mode and its
frequency at $\Phi_{C}(0)$) due to the Coulomb interactions is approximately
$\sim0.3\Delta f$ which is compatible with the small value of $\kappa\approx0.2\times(unit\,cell\,length)^{-1}$. 

The resonance frequencies depend on the flux periodically with the
period $\Delta\Phi=\Phi_{0}$. We have observed the mode softening
with approaching the quantum critical points, symmetrically positioned
with respect to the flux $\Phi_{0}/2$. This is the key feature of
the transverse field Ising model. According to this model, the phases
on both sides of the quantum critical point (the quantum paramagnet
and ferromagnet, in the Ising spin terminology) are characterized
by different types of excitations. In the paramagnetic phase the relevant
excitations are flips of individual spins. These excitations become
gapless exactly at the critical point. In contrast, the excitations
in the ordered (ferromagnetic) phase are the domain walls (or described
as kinks) between different ground states (\textquotedblleft vacua\textquotedblright ).
Even though any superconductor is a bosonic system, the lowest-energy
excitations in the ladders are expected to have fermionic nature;
the higher-energy excitations are composite particles made of these
fermions (similar to the appearance of quarks and mesons). The emergence
of non-trivial excitations near the quantum phase transition and their
properties is one of the main themes of the present and future research. 

The gradual shift of the minima of different modes shown in Fig. \ref{MW data on chain spectra}b
is due to a non-zero ground capacitance of the unit cells that result
in weak but non-negligible long-range interactions between the pseudo-spins
in agreement with the theory presented in Section III. The non-zero
$C_{G}$ is not a ``bug'' but a ``feature'': variation of the
ground capacitance of individual cells enables fabrication of the
devices characterized by different types of coupling between the unit
cells, which makes possible emulation of different quantum models
such as Ising and XY models in transverse magnetic fields. Interactions
between the cells of Josephson arrays (\textquotedblleft spins\textquotedblright )
can be tuned by changing the Josephson energies of the larger and
smaller junctions. This tuning remains a challenge for other solid-state
systems \cite{Dutta2015,Sachdev1999}. 

The accuracy of tracing the low-energy modes in the ``ferromagnetic''
phase is much lower than that in the ``paramagnetic'' phase (see
Fig. 6b). Figure 7 shows that in the single-tone measurements with
a long 92-unit-cell chain, the readout resonance is dramatically broadened
between the critical points. For this reason, in the two-tone measurements
we could not accurately determine the positions of low-energy modes
for this chain in the ferromagnetic phase. We attribute this resonance
``smearing'' to a strong magnetic (flux) noise in our system. As
shown in section \ref{subsec:Matrix-elements-of}, in the ordered
(ferromagnetic) phase the effect of the charge noise on the transitions
between the low-energy states is weaker, whereas the effect of the
flux noise is much stronger than that in the disordered (paramagnetic)
phase. Note that the decay is proportional to the square of the matrix
elements shown in Fig. \ref{fig:Matrix-elements-of}, so the decay
due to the flux noise is enhanced by a factor of $\sim10^{6}$ in
the ordered phase while the decay due to the charge noise is reduced
by a factor of $\sim10^{3}$. The effect of the charge noise is further
suppressed by a small factor that translates the matrix element of
the Ising operators into the matrix element of the physical charge
(the effect is opposite for the flux noise). By comparing equations
(\ref{eq:J_eff},\ref{eq:Q_eff}) relating the dimensionless matrix
operators shown in Fig. \ref{fig:Matrix-elements-of} to the physical
noise, we see that the decay rate due to the charge noise is additionally
decreased by a factor $(t/E_{c}\eta_{0})^{2}\lesssim\delta r_{c}$
whereas the decay rate of the flux noise is enhanced by the factor
$(E_{JL}\eta_{0}/\omega)^{2}$ where $\omega$ is the mode frequency.
This makes the effect of the flux noise on the ordered side of the
transition very large whilst the effect of the charge noise remains
moderate. 

\begin{figure}
\includegraphics[width=1\columnwidth]{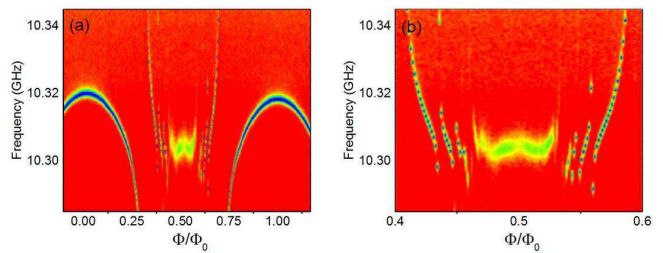}\caption{The single-tone measurements for a long (92-unit-cell) chain {[}panel
(b) is the blow-up of the data within the range $0.4<\Phi/\Phi_{0}<0.6${]}.
The avoided crossings between the low-energy modes and the readout
resonator mode are clearly observed in the ``paramagnetic'' phase
($\Phi/\Phi_{0}<0.45,\Phi/\Phi_{0}>0.55$), where the readout resonance
is sharp (the resonance width $\Delta f<1MHz$). In the ``ferromagnetic''
phase ($0.45<\Phi/\Phi_{0}<0.55$) the resonance is smeared ($\Delta f\approx4MHz$)
due to transitions between the low-energy modes induced by the flux
noise (see the text for more details). \label{resonance smearing}}
\end{figure}

\section{Conclusion and Outlook}

Controllable-by-design interactions between the pseudospins implemented
as two states of the unit cells of periodic Josephson ladders enabled
us to study the phase boundaries between the ordered and disordered
phases and critical behavior close to this transition. In particular,
the study of microwave properties allowed us to characterize the low-energy
spectrum of this system. 

Close to the transition the low-energy degrees of this system are
described by the $\varphi^{4}$ model \cite{Sachdev1999} that displays
the transition between the ordered and disordered phases that belongs
to the Ising universality class. In the Ising model the low energy
excitations are Majorana quasiparticles, thus our observation of the
low energy modes of the long ladders can be viewed as a direct probe
of Majorana excitations in this system. Note that, in contrast to
the spin chains where only pairs of Majorana particles can be produced
and studied by scattering techniques, Josephson ladders will allow
us to sudy the full low energy spectrum, in particular the fermionic
excitations. An appealing analogy is provided by the particle physics
in which fermionic excitations are quarks while pair of those correspond
to pions. In this analogy the spectrum shown in Fig. \ref{MW data on chain spectra}
corresponds to quarks while the spectrum studied in spin chains -
to pions. 

The developed experimental platform can help to address various fundamental
issues. First, disorder can be controllably introduced in this 1D
system, and this enables the study of the appearance of intermediate
glassy non-ergodic phases. Second, the platform allows for engineering
interactions between pseudospins that violate exact integrability
of the effective Ising model that describes the system at large scales.
For instance, the ground state capacitance results in long-range attractions
between kinks in the Ising model in the ordered state. This allows
one to study how ergodic behavior reappears in almost integrable quantum
systems. 

We envision the future work that will address the emergence of the
symmetries near the critical point in the Josephson ladders. By measuring
microwave responses of the ladders in different phases and extracting
the excitation spectra and their matrix elements, one can fully explore
the behavior of this system near the critical point and extract the
critical indices. By introducing controllable scattering in the junction
parameters, one can explore the effects of disorder on the quantum
phase transitions in 1D. 

Another exciting line of research is the effect of non-integrability
on the excitation decay and dephasing. Because in the critical regime
the lowest energy excitations are equivalent to the ones of the effective
Ising model at the critical point, the integrability implies that
these excitations have an infinite lifetime. In realistic arrays the
integrability becomes approximate and excitations acquire a finite,
albeit long lifetime. In the extensive time-domain measurements, the
decoherence rate in the 1D Josephson ladders can be extracted from
Rabi oscillations and Ramsey fringes as a function of proximity to
the critical point. We expect that the residual interaction between
quasiparticles leads to much faster decoherence rates at higher energy.
These measurements address an important issue of the emergence of
classicality in closed quantum systems. 

A potential extension of these experiments is the intentional gradual
variation of the unit cell parameters in the ladders. The idea is
to produce ladders characterized by a smooth transition region between
the states with broken and unbroken symmetry. The position of the
phase boundary (or distance between two boundaries) within a ladder
with gradual change in the position-dependent parameter $r(x)$ can
be varied by the magnetic field that controls the proximity of individual
cells to the critical point. We expect that the boundary between single-well
(\textquotedblleft paramagnetic\textquotedblright ) and double-well
(\textquotedblleft ferromagnetic\textquotedblright ) phases supports
the zero-energy (Majorana) mode that can be probed by spectroscopic
methods. A crucial question that we plan to address is the mechanisms
of the decoherence relevant for this mode.

Finally, by combining few ladders together one can get the experimental
realization of very exotic quantum objects such as the two channel
Kondo problem with its $1/2ln2$ entropy due to exactly degenerate
Majorana quasipartciles \cite{pino2015}. 

\section{Acknowledgements}

The work at Rutgers was supported by the NSF award 1708954, ARO grant
W911NF-13-1-0431. The work at the University of Massachusetts Boston
was supported in part by NSF awards ECCS-1608448 and DUE-1723511.

\bibliographystyle{plain}
\bibliography{1DQPT}

\end{document}